\documentclass[preprints,article,accept,moreauthors,pdftex]{mdpi}

\usepackage{subfig}
\usepackage{upgreek}
\usepackage{units}
\usepackage{amssymb}
\usepackage{tabularx}
\usepackage{todonotes}
\usepackage{dsfont}
\usepackage{cancel}
\graphicspath{{./Figures/pdf/}}

\firstpage{1}
\pubvolume{xx}
\issuenum{1}
\articlenumber{5}
\pubyear{2019}
\copyrightyear{2019}
\history{Received: 31 August 2019; Accepted: 17 September 2019; Published: 24 September 2019}

\renewcommand{\eqref}[1]{\mbox{Equation~(\ref{#1})}}

\newcommand{\secref}[1]{\mbox{Section~\ref{#1}}}

\newcommand{\appref}[1]{\mbox{Appendix~\ref{#1}}}

\updates{yes} 

\Title{Formal Developments for Lorentz-Violating Dirac Fermions and Neutrinos}

\Author{Jo\~{a}o Alf\'{i}eres Andrade de Sim\~{o}es dos Reis $^{1,2}$\orcidA{}, Marco Schreck $^{1,}$*\orcidB{}}
\AuthorNames{Jo\~{a}o Alf\'{i}eres Andrade de Sim\~{o}es dos Reis and Marco Schreck}

\address{$^{1}$ \quad Departamento de F\'{\i}sica, Universidade Federal do Maranh\~{a}o, Campus Universit\'{a}rio do Bacanga, S\~{a}o Lu\'{\i}s--MA 65080-805, Brazil; jalfieres@gmail.com (J.A.A.d.S.d.R.) \\ 
$^{2}$ \quad Departamento de F\'{i}sica, Centro de Educa\c{c}\~{a}o, Ci\^{e}ncias Exatas e Naturais, Universidade Estadual do Maranh\~{a}o, Cidade Universit\'{a}ria Paulo VI, S\~{a}o Lu\'{i}s--MA 65055-310, Brazil}

\corres{Correspondence: marco.schreck@ufma.br; Tel.: +55-(98)3272-8293}

\abstract{The current paper is a technical work that is focused on Lorentz violation for Dirac fermions as well as neutrinos, described within the nonminimal Standard-Model Extension. We intend to derive two theoretical results. The first is the full propagator of the single-fermion Dirac theory modified by Lorentz violation. The second is the dispersion equation for a theory of $N$ neutrino flavors that enables the description of both Dirac and Majorana neutrinos. As the matrix structure of the neutrino field operator is very involved for generic $N$, we will use sophisticated methods of linear algebra to achieve our objectives. Our main finding is that the neutrino dispersion equation has the same structure in terms of Lorentz-violating operators as that of a modified single-fermion Dirac theory. The results will be valuable for phenomenological studies of Lorentz-violating Dirac fermions and~neutrinos.}


\begin{document}


\section{Introduction}

Neutrinos are both interesting and elusive particles. According to the Standard Model of particle physics, there are three neutrino flavors that correspond to the flavors of the three charged leptons—the electron neutrino $\upnu_{\mathrm{e}}$, the muon neutrino $\upnu_{\upmu}$ and the tau neutrino $\upnu_{\uptau}$. Neutrinos do not carry electric charge and are only produced in processes mediated by the weak interaction \cite{Tanabashi:2018oca}. When neutrinos propagate a distance, the probability of detecting a certain flavor changes with time. These neutrino oscillations are quantum mechanical in nature. They have their origin in the fact that the eigenstates of the kinematic Hamiltonian and the flavor eigenstates produced in interactions do not correspond to each other. Instead, these two distinct bases are related by the unitary PMNS 
 matrix \cite{Bilenky:2004xm}. Neutrino oscillations indicate that neutrinos have mass, although their mass is that tiny to not have been measured directly, so far. Therefore, they practically propagate with the speed of light.

Measuring the {\em CP}-violating phase that is contained in the PMNS matrix is currently one of the hot topics in neutrino physics. Apart from {\em CP}-violation, neutrinos may be subject to \emph {CPT} violation due to physics at the Planck scale such as strings \cite{Kostelecky:1988zi,Kostelecky:1995qk}. Since they travel almost with the speed of light, they are strongly boosted with respect to the Sun-centered equatorial frame \cite{Kostelecky:2008ts} and {\em CPT}-violating effects may accumulate over long distances. A violation of \emph{CPT} symmetry implies a violation of Lorentz invariance in the context of effective field theory \cite{Greenberg:2002uu}, whereby it makes sense to use neutrinos as a testbed for the search for Lorentz violation. To do so, a general comprehensive framework to parameterize Lorentz violation is desirable. The latter is provided by the Standard-Model Extension (SME). Its minimal version including field operators of mass dimensions 3 and 4 was developed in References \cite{Colladay:1998fq,Colladay:1996iz}. The nonminimal SME, which involves field operators of arbitrary mass dimensions, was constructed in a series of papers \cite{Kostelecky:2009zp,Kostelecky:2011gq,Kostelecky:2013rta} for photons, neutrinos and single Dirac fermions. Within this framework, each Lorentz-violating contribution is a proper contraction of a field operator and a background field. The latter are composed of preferred spacetime directions and controlling coefficients parameterizing the strength of Lorentz violation.

Phenomenology in the SME neutrino sector has been performed in a large collection of papers~ \mbox{\cite{Katori:2006mz,Diaz:2009qk,Katori:2010nf,Diaz:2010ft,AguilarArevalo:2011yi,Diaz:2011ia,Katori:2011zz,Abe:2012gw,Katori:2012pe,Diaz:2013saa,Diaz:2013wia,Diaz:2013ywa,Diaz:2014yva,Diaz:2014hca,Diaz:2015dxa,Diaz:2015aua,Diaz:2016fqd,Diaz:2016xpw,Arguelles:2016rkg,Katori:2016eni,Abe:2017eot,Aartsen:2017ibm}}, where this list is not claimed to be complete. All constraints obtained for Lorentz violation in the neutrino sector are compiled in the data tables~\cite{Kostelecky:2008ts}. Interestingly, it was also observed that neutrino oscillations could be explained by Lorentz-violating \emph{massless} neutrinos in certain models.

The current article must be considered as a technical work that can be the preparing base for further forthcoming phenomenological investigations. We have two objectives. The first is to present the general result for the propagator of a single Dirac fermion in the nonminimal SME. The second is to obtain the dispersion equation of the nonminimal neutrino sector. The latter goal will be accomplished with a powerful method to compute determinants of large matrices that decompose into smaller~blocks.

A description of Lorentz violation in the neutrino sector based on the SME permits $N$ neutrino flavors and allows for both Dirac and Majorana neutrinos. Majorana neutrinos are characterized by the property of being identical to their own antiparticles such that neutrino-antineutrino mixing can occur. Therefore, the differential operator that appears in the field equations of the theory is not simply a $(4\times 4)$ matrix in spinor space such as for a single Dirac fermion. For a set of $N$ neutrinos of either Dirac or Majorana type, it is a $(8N\times 8N)$ matrix instead. The determinant of this matrix directly corresponds to the modified dispersion equation for neutrinos. Even for 3 flavors, its computation is~ cumbersome.

In principle, it is possible to consider an observer frame with only a single nonzero coefficient and to compute the determinant by brute force with computer algebra. Such a direct approach has several disadvantages, though. First, an observer frame with a single nonzero controlling coefficient is a very special case. Second, the result of the determinant is most probably still messy and its structure is supposedly not very illuminating. Therefore, it would be desirable to employ a technique that allows for a covariant and general treatment of the problem. Although the result is still expected to be complicated due to the high dimensionality of the matrix, the method to be used can be applied to obtain the dispersion equation for an arbitrary number of flavors. Thus, it may be of interest for someone who wants to include sterile neutrinos in their analysis, which are beyond the scope of the~SME.

The paper is organized as follows. Section~\ref{sec:basic-properties-fermions} gives a brief introduction to the SME fermion sector. The properties most important to us are discussed and several definitions are introduced. In \secref{sec:dirac-propagator} we derive the full propagator of the modified Dirac fermion sector. In \secref{sec:neutrino-dispersion-equation-computation} the very base of the SME neutrino sector is described, as well as the algorithm that we intend to use to obtain the neutrino dispersion equation. The individual steps of the calculation are carried out and explained, too. We state the central result in \secref{sec:neutrino-dispersion-equation-result} and discuss it subsequently, whereby some properties of the first-order dispersion relations are obtained in \secref{sec:first-order-dispersion-relations}. A brief comment on classical Lagrangians in the neutrino sector follows in \secref{sec:classical-lagrangians}. Finally, all findings are summarized and concluded on in \secref{sec:conclusions}. Relations and definitions that are not of primary interest to the reader are relegated to Appendixes~\ref{sec:relations-dirac-matrices} and \ref{sec:definition-wedge-hodge}. Natural units with the conventions $\hbar=c=1$ will be used unless otherwise~ stated.

\section{Basic Properties of the SME Fermion Sector}
\label{sec:basic-properties-fermions}

The fermion sector of the SME describes a single Dirac fermion subject to Lorentz-violating background fields that permit a construction of an observer Lorentz-invariant Lagrange density. The minimal fermion sector was introduced in Reference \cite{Colladay:1998fq} and its properties were investigated in Reference \cite{Kostelecky:2000mm}. The minimal framework was complemented by the nonminimal contributions in Reference \cite{Kostelecky:2013rta}. Our analysis will be carried out within the nonminimal SME, whereby we take over the notation of the latter reference and also mainly refer to formulas stated in that paper.

The framework rests on the Lagrange density of Equations~(1) and (2) in \cite{Kostelecky:2013rta}. The corresponding modified Dirac equation reads
\begin{subequations}
\label{eq:single-fermion-modified-dirac-theory}
\begin{align}
\mathcal{D}\psi&=0\,,\quad \mathcal{D}=\cancel{p}-m_{\psi}\mathds{1}_4+\hat{\mathcal{Q}}\,, \displaybreak[0]\\[2ex]
\hat{\mathcal{Q}}&=\hat{\mathcal{S}}\mathds{1}_4+\mathrm{i}\hat{\mathcal{P}}\gamma^5+\hat{\mathcal{V}}^{\mu}\gamma_{\mu}+\hat{\mathcal{A}}^{\mu}\gamma^5\gamma_{\mu}+\frac{1}{2}\hat{\mathcal{T}}^{\mu\nu}\sigma_{\mu\nu}\,.
\end{align}
\end{subequations}
In the latter, $\psi$ is a spinor, $m_{\psi}$ the fermion mass, and $\gamma^{\mu}$ denote the usual Dirac matrices that satisfy the Clifford algebra $\{\gamma^{\mu},\gamma^{\nu}\}=2\eta^{\mu\nu}\mathds{1}_4$ with the Minkowski metric $\eta_{\mu\nu}$ of signature $(+,-,-,-)$. Furthermore, $\mathds{1}_n$ is the $n$-dimensional identity matrix, $\gamma_5=\gamma^5\equiv\mathrm{i}\gamma^0\gamma^1\gamma^2\gamma^3$ is the chiral Dirac matrix and $\sigma_{\mu\nu}\equiv (\mathrm{i}/2)[\gamma_{\mu},\gamma_{\nu}]$ involves the commutator of two Dirac matrices. The Lorentz-violating operator $\hat{\mathcal{Q}}$ is decomposed in terms of the 16 matrices $\{\Gamma^A\}\equiv \{\mathds{1}_4,\gamma^5,\gamma^{\mu},\mathrm{i}\gamma^5\gamma^{\mu},\sigma^{\mu\nu}\}$. This set forms a basis of $(4\times 4)$ matrices and the dual basis $\{\Gamma_A\}$ is obtained by lowering the Lorentz indices with the Minkowski metric. The basis obeys an orthogonality relation of the form $\mathrm{Tr}(\Gamma_A\Gamma^B)=4\delta_A^{\phantom{A}B}$ where $\mathrm{Tr}$ denotes the trace in spinor space.
Lorentz violation is contained in a scalar $\hat{\mathcal{S}}$, a vector $\hat{\mathcal{V}}$, and a two-tensor operator $\hat{\mathcal{T}}$. Additionally, a pseudo-scalar $\hat{\mathcal{P}}$ and a pseudo-vector $\hat{\mathcal{A}}$ occur when the behavior of the operators under parity transformations is taken into account. Tensors of higher rank than these do not exist, as more complicated matrices in spinor space can always be mapped in some way to the 16 matrices mentioned before by using identities such as those listed in \appref{sec:relations-dirac-matrices}.

It is worth pointing out the structure of the operators contained in $\hat{\mathcal{Q}}$. They are constructed as sums of operators of increasing mass dimension suitably contracted with controlling coefficients. In momentum space they can be written as the following infinite sums:
\begin{subequations}
\label{eq:operators}
\begin{align}
\hat{\mathcal{S}}=&\sum_{d=3}^{\infty}\mathcal{S}^{(d)\alpha_{1}\alpha_{2}\ldots\alpha_{d-3}}p_{\alpha_{1}}p_{\alpha_{2}}\ldots p_{\alpha_{d-3}}\,, \displaybreak[0]\\[2ex]
\hat{\mathcal{P}}=&\sum_{d=3}^{\infty}\mathcal{P}^{(d)\alpha_{1}\alpha_{2}\ldots\alpha_{d-3}}p_{\alpha_{1}}p_{\alpha_{2}}\ldots p_{\alpha_{d-3}}\,, \displaybreak[0]\\[2ex]
\hat{\mathcal{V}}^{\mu}=&\sum_{d=3}^{\infty}\mathcal{V}^{(d)\mu\alpha_{1}\alpha_{2}\ldots\alpha_{d-3}}p_{\alpha_{1}}p_{\alpha_{2}}\ldots p_{\alpha_{d-3}}\,, \displaybreak[0]\\[2ex]
\hat{\mathcal{A}}^{\mu}=&\sum_{d=3}^{\infty}\mathcal{A}^{(d)\mu\alpha_{1}\alpha_{2}\ldots\alpha_{d-3}}p_{\alpha_{1}}p_{\alpha_{2}}\ldots p_{\alpha_{d-3}}\,, \displaybreak[0]\\[2ex]
\hat{\mathcal{T}}^{\mu\nu}=&\sum_{d=3}^{\infty}\mathcal{T}^{(d)\mu\nu\alpha_{1}\alpha_{2}\ldots\alpha_{d-3}}p_{\alpha_{1}}p_{\alpha_{2}}\ldots p_{\alpha_{d-3}}\,,
\end{align}
\end{subequations}
where $d$ is the mass dimension of the field operator that a specific controlling coefficient such as $\mathcal{S}^{(d)\alpha_{1}\ldots\alpha_{d-3}}$ is contracted with. These decompositions can be extracted from Equation~(3) in \cite{Kostelecky:2013rta}. Each controlling coefficient has a mass dimension of $4-d$ and is independent of the spacetime coordinates to preserve energy and momentum.

Evaluating the determinant of the Dirac operator $\mathcal{D}$ leads to the dispersion equation for a single fermion. The latter is given  by Equation~(39) of \cite{Kostelecky:2013rta}:
\begin{subequations}
\label{eq:dispersion-equation-single-fermion}
\begin{equation}
\Delta=0\,,\quad \Delta=(\hat{\mathcal{S}}_-^2-\hat{\mathcal{T}}_-^2)(\hat{\mathcal{S}}_+^2-\hat{\mathcal{T}}_+^2)+\hat{\mathcal{V}}_-^2\hat{\mathcal{V}}_+^2-2\hat{\mathcal{V}}_-\cdot(\hat{\mathcal{S}}_-\eta+2\mathrm{i}\hat{\mathcal{T}}_-)\cdot (\hat{\mathcal{S}}_+\eta-2\mathrm{i}\hat{\mathcal{T}}_+)\cdot\hat{\mathcal{V}}_+\,,
\end{equation}
with the definitions
\begin{equation}
\hat{\mathcal{S}}_{\pm}\equiv-m_{\psi}+\hat{\mathcal{S}}\pm\mathrm{i}\hat{\mathcal{P}}\,,\quad \hat{\mathcal{V}}_{\pm}^{\mu}\equiv p^{\mu}+\hat{\mathcal{V}}^{\mu}\pm\hat{\mathcal{A}}^{\mu}\,,\quad \hat{\mathcal{T}}^{\mu\nu}_{\pm}\equiv\frac{1}{2}(\hat{\mathcal{T}}^{\mu\nu}\pm\mathrm{i}\widetilde{\hat{\mathcal{T}}}{}^{\mu\nu})\,,
\end{equation}
and the dual of the two-tensor operator:
\begin{equation}
\widetilde{\mathcal{\hat{T}}}{}^{\mu\nu}\equiv\frac{1}{2}\varepsilon^{\mu\nu\varrho\sigma}\hat{\mathcal{T}}_{\varrho\sigma}\,.
\end{equation}
\end{subequations}

The latter contains the totally antisymmetric Levi-Civita symbol $\varepsilon^{\mu\nu\varrho\sigma}$  based on the convention $\varepsilon^{0123}=1$. In the remainder of the paper, a definition of the observer scalars
\begin{equation}
\hat{X}\equiv \frac{1}{4}\hat{\mathcal{T}}_{\mu\nu}\hat{\mathcal{T}}^{\mu\nu}\,,\quad \hat{Y}\equiv \frac{1}{4}\hat{\mathcal{T}}_{\mu\nu}\widetilde{\mathcal{\hat{T}}}{}^{\mu\nu}\,,
\end{equation}
turns out to be fruitful. It is also beneficial to define the following combination of operators that we will make frequent use of:
\begin{equation}
\widetilde{\mathcal{\hat{T}}}{}_{\mathrm{gen}}^{\mu\nu}\equiv \widetilde{\mathcal{\hat{T}}}{}^{\mu\nu}-\frac{1}{\hat{\mathcal{S}}-m_{\psi}}\left[(p+\hat{\mathcal{V}})^{\mu}\hat{\mathcal{A}}^{\nu}-\hat{\mathcal{A}}^{\mu}(p+\hat{\mathcal{V}})^{\nu}\right]\,,\quad \hat{\mathcal{T}}_{\mathrm{gen}}^{\mu\nu}\equiv\frac{1}{2}\varepsilon^{\mu\nu\varrho\sigma}(\widetilde{\mathcal{\hat{T}}}_{\mathrm{gen}})_{\varrho\sigma}\,.
\end{equation}

Note that the operator previously introduced reduces to the effective dual two-tensor operator in~Equation~(25) of Reference~\cite{Kostelecky:2013rta} at first order in Lorentz violation, which is why we denote it by the index ``gen'' standing for ``generalized.''

\section{General Modified Dirac Propagator}
\label{sec:dirac-propagator}

The investigations to be performed in the neutrino sector require an evaluation of the single-fermion propagator (Green's function in momentum space) $S$. This result has not been obtained so far for the full nonminimal fermion sector, which is why we would like to state it here. The propagator is directly connected to the inverse of the Dirac operator in momentum space: $\mathcal{D}S=S\mathcal{D}=\mathds{1}_4$. It must be possible to express the propagator in terms of the basis $\{\Gamma^A\}$ mentioned before. Therefore, it can be written in the form
\begin{equation}
\label{eq:propagator-single-fermion}
\mathrm{i}S=\frac{\mathrm{i}}{\Delta}\left(\hat{S}^{(p)}\mathds{1}_{4}+\mathrm{i}\hat{\mathcal{P}}^{(p)}\gamma^{5}+\hat{\mathcal{V}}^{(p)\mu}\gamma_{\mu}+\hat{\mathcal{A}}^{(p)\mu}\gamma^{5}\gamma_{\mu}+\frac{1}{2}\hat{\mathcal{T}}^{(p)\mu\nu}\sigma_{\mu\nu}\right)\,,
\end{equation}
where $\Delta=\det(\mathcal{D})$ and the index $(p)$ of each individual contribution stands for ``propagator.'' Note that we introduced a prefactor of $\mathrm{i}$ to follow the conventions of Reference \cite{Peskin:1995}. The denominator $\Delta$ corresponds to the left-hand side of \eqref{eq:dispersion-equation-single-fermion}. The individual contributions can be obtained by multiplying the inverse with each of the 16 Dirac matrices and computing the trace of the matrix product. In addition, we make use of the orthogonality relation for these matrices, which leads~to:
\begin{subequations}
\begin{align}
\hat{\mathcal{S}}^{(p)}&=\frac{\Delta}{4}\mathrm{Tr}(\mathds{1}_4\mathcal{D}^{-1})\,,\quad \hat{\mathcal{P}}^{(p)}=-\mathrm{i}\frac{\Delta}{4}\mathrm{Tr}(\gamma^5\mathcal{D}^{-1})\,,\quad \hat{\mathcal{V}}^{(p)\mu}=\frac{\Delta}{4}\mathrm{Tr}(\gamma^{\mu}\mathcal{D}^{-1})\,, \displaybreak[0]\\[2ex]
\hat{\mathcal{A}}^{(p)\mu}&=-\frac{\Delta}{4}\mathrm{Tr}(\gamma^5\gamma^{\mu}\mathcal{D}^{-1})\,,\quad \hat{\mathcal{T}}^{(p)\mu\nu}=\frac{\Delta}{4}\mathrm{Tr}(\sigma^{\mu\nu}\mathcal{D}^{-1})\,.
\end{align}
\end{subequations}

Now, these contributions are explicitly given by
\begin{subequations}
\label{eq:propagator-contributions}
\begin{align}
\hat{\mathcal{S}}^{(p)}&=-(\hat{\mathcal{S}}-m_{\psi})(2\Theta -\hat{\mathcal{T}}_{\mu\nu}\hat{\mathcal{T}}_{\mathrm{gen}}^{\mu\nu})-2\hat{Y}\hat{\mathcal{P}}\,, \displaybreak[0]\\[2ex]
\hat{\mathcal{P}}^{(p)}&=-(\hat{\mathcal{S}}-m_{\psi})(2\hat{Y}-\hat{\mathcal{T}}_{\mu\nu}\widetilde{\mathcal{\hat{T}}}{}_{\mathrm{gen}}^{\mu\nu})+2\Theta \hat{\mathcal{P}}\,, \displaybreak[0]\\[2ex]
\hat{\mathcal{V}}^{(p)\mu}&=2\left[\Theta(p+\hat{\mathcal{V}})^{\mu}-\hat{\mathcal{T}}^{\mu \nu }\hat{\mathcal{T}}_{\nu\varrho}(p+\hat{\mathcal{V}})^{\varrho}-(\hat{\mathcal{S}}-m_{\psi})\widetilde{\mathcal{\hat{T}}}{}_{\mathrm{gen}}^{\mu \nu}\hat{\mathcal{A}}_{\nu}+\hat{\mathcal{T}}^{\mu\nu}\hat{\mathcal{A}}_{\nu}\hat{\mathcal{P}}\right]\,, \displaybreak[0]\\[2ex]
\hat{\mathcal{A}}^{(p)\mu}&=2\left[\Theta\hat{\mathcal{A}}^{\mu}-\hat{\mathcal{T}}^{\mu\nu}\hat{\mathcal{T}}_{\nu\varrho}\hat{\mathcal{A}}^{\varrho}-(\hat{\mathcal{S}}-m_{\psi})\widetilde{\mathcal{\hat{T}}}{}_{\mathrm{gen}}^{\mu\nu}(p+\hat{\mathcal{V}}) _{\nu}+\hat{\mathcal{T}}^{\mu\nu}(p+\hat{\mathcal{V}})_{\nu}\hat{\mathcal{P}}\right]\,, \displaybreak[0]\\[2ex]
\hat{\mathcal{T}}^{(p)\mu\nu}&=2\left[(\hat{\mathcal{S}}-m_{\psi})^{2}\hat{\mathcal{T}}{}_{\mathrm{gen}}^{\mu\nu}-\Theta\hat{\mathcal{T}}^{\mu\nu}+\hat{Y}\widetilde{\mathcal{\hat{T}}}{}^{\mu\nu}+\left[\hat{\mathcal{T}}^{\mu\varrho}(p+\hat{\mathcal{V}})^{\nu}-(p+\hat{\mathcal{V}}) ^{\mu}\hat{\mathcal{T}}^{\nu\varrho}\right](p+\hat{\mathcal{V}})_{\varrho}\right. \notag \\
&\phantom{{}={}2[}\left.-(\hat{\mathcal{T}}^{\mu \varrho}\hat{\mathcal{A}}^{\nu}-\hat{\mathcal{A}}^{\mu}\hat{\mathcal{T}}^{\nu \varrho})\hat{\mathcal{A}}_{\varrho}-( \hat{\mathcal{S}}-m_{\psi})\hat{\mathcal{P}}\widetilde{\mathcal{\hat{T}}}{}_{\mathrm{gen}}^{\mu\nu}\right]\,.
\end{align}

For convenience, we defined the observer scalar
\begin{equation}
2\Theta\equiv(p+\hat{\mathcal{V}})^{2}-(\hat{\mathcal{S}}-m_{\psi})^{2}-\hat{\mathcal{A}}^{2}-2\hat{X}-\hat{\mathcal{P}}^{2}\,,
\end{equation}
\end{subequations}
which involves each of the five operators. Several remarks are in order. First, this result generalizes the propagator obtained for the spin-degenerate operators $\hat{\mathcal{S}}$, $\hat{\mathcal{V}}$ in Reference \cite{Schreck:2017isa} and that for the spin-nondegenerate operators $\hat{\mathcal{A}}$, $\hat{\mathcal{T}}$ in Reference \cite{Reis:2016hzu}. It now applies to the full spectrum of Lorentz-violating operators and is valid also for the nonminimal SME. The propagator reduces to the special results published previously when the corresponding operators are set to zero. Second, all operators of different types are coupled to each other and each contribution of Equation~(\ref{eq:propagator-contributions}) transforms consistently under parity transformations, as expected. For example, each term of $\hat{\mathcal{P}}^{(p)}$ transforms as a pseudoscalar. Third, for vanishing Lorentz violation, we have
\begin{subequations}
\begin{align}
2\Theta&=p^2-m_{\psi}^2\,,\quad \hat{\mathcal{S}}^{(p)}=m_{\psi}(p^2-m_{\psi}^2)\,,\quad \hat{\mathcal{P}}^{(p)}=0\,, \displaybreak[0]\\[2ex]
\hat{\mathcal{V}}^{(p)\mu}&=(p^2-m_{\psi}^2)p^{\mu}\,,\quad \hat{\mathcal{A}}^{(p)\mu}=0\,,\quad \hat{\mathcal{T}}^{(p)\mu\nu}=0\,,\quad \Delta=(p^2-m_{\psi}^2)^2\,,
\end{align}
\end{subequations}
whereupon \eqref{eq:propagator-single-fermion} reproduces the standard fermion propagator
\begin{equation}
\mathrm{i}S|_{\text{LV}=0}=\frac{\mathrm{i}(\cancel{p}+m_{\psi}\mathds{1}_4)}{p^2-m_{\psi}^2}\,,
\end{equation}
stated in \cite{Peskin:1995}.

\section{Modified Neutrino Dispersion Equation}
\label{sec:neutrino-dispersion-equation-computation}

We consider $N$ flavors of modified neutrinos and also include a description of Majorana neutrinos. To do so, the spinor field $\Psi_A$ is constructed as a $2N$-dimensional multiplet of spinors
\begin{equation}
\label{eq:spinorfield}
\Psi_A=\begin{pmatrix}
\psi_a \\
\psi_a^C \\
\end{pmatrix}\,,
\end{equation}
where $a$ ranges over $N$ flavors and $A$ labels the $2N$ components of the multiplet. Furthermore, $\psi_a^C$ is the charge conjugate of $\psi_a$ \cite{Kostelecky:2011gq}. Due to the form of the construction, there is a redundancy in $\Psi$ encoded in the relationship
\begin{equation}
\label{eq:relationshipPsi}
\Psi^C=\mathcal{C}\Psi\,,\quad \mathcal{C}=\begin{pmatrix}
0 & \mathds{1}_N \\
\mathds{1}_N & 0 \\
\end{pmatrix}\,,
\end{equation}
where the $(2N\times 2N)$ matrix $\mathcal{C}$ is defined in terms of $(N\times N)$ blocks in flavor space. With this information at hand, we present the Lagrange density that incorporates Lorentz and \emph{CPT} violation into the neutrino sector:
\begin{equation}
\label{eq:LagrangeDensity}
\mathcal{L}=\frac{1}{2}\overline{\Psi}_{A}(\mathrm{i}\cancel{\partial}\delta_{AB}-M_{AB}+\hat{\mathcal{Q}}_{AB})\Psi_{B}+\mathrm{H.c.}\,,
\end{equation}
with the flavor indices $A$, $B$. The corresponding Dirac operator $\mathcal{D}_{\upnu}$ in momentum space is a $(8N\times 8N)$ matrix that can be expressed in the form \cite{Kostelecky:2011gq}
\begin{subequations}
\begin{equation}
\label{eq:dirac-operator-modified-neutrinos}
\mathcal{D}_{\upnu}=\mathds{1}_{2N}\otimes\cancel{p}-M\otimes\mathds{1}_{4}+\mathcal{\hat{Q}}\,,
\end{equation}
where
\begin{equation}
\mathcal{\hat{Q}}=\mathcal{\hat{S}}\otimes\mathds{1}_4+\mathrm{i}\mathcal{\hat{P}}\otimes\gamma^{5}+\mathcal{\hat{V}}^{\mu}\otimes\gamma_{\mu}+\mathcal{\hat{A}}^{\mu}\otimes\gamma ^{5}\gamma_{\mu}+\frac{\mathrm{i}}{2}\mathcal{\hat{T}}^{\mu\nu}\otimes\sigma_{\mu\nu}\,.
\end{equation}
\end{subequations}

Here, $\otimes$ denotes a tensor product of $(2N\times 2N)$ matrices in flavor space and matrices in four-dimensional spinor space. Furthermore, $M$ is the $(2N\times 2N)$ neutrino mass matrix. Both the mass matrix and the Lorentz-violating operator $\hat{\mathcal{Q}}$ are expressed in the basis of flavor eigenstates. Therefore, the mass matrix cannot simply be taken as diagonal. The operator $\hat{\mathcal{Q}}$ can be decomposed as in~Equation~(\ref{eq:operators}), but now we need to remember that the individual contributions also have a flavor~structure.

The dispersion equation corresponds to the determinant of the Dirac operator set equal to zero. Therefore, the basic problem is to obtain this determinant and to express the result in a convenient manner. It turns out to be very useful to treat the Dirac operator as a $(2N\times 2N)$ matrix in flavor space where each of these entries on its own is a $(4\times 4)$ matrix in spinor space. Thus, the structure of this operator is
\begin{subequations}
\begin{equation}
\label{eq:dirac-operator-neutrinos}
\mathcal{D}_{\upnu}=\begin{pmatrix}
\mathcal{D}_{1,1} & \dots & \mathcal{D}_{1,2N} \\
\vdots & \ddots & \vdots \\
\mathcal{D}_{2N,1} & \dots & \mathcal{D}_{2N,2N} \\
\end{pmatrix}\,,
\end{equation}
where
\begin{align}
\mathcal{D}_{i,j}&=\delta_{i,j}\cancel{p}-M_{i,j}\mathds{1}_4+\hat{\mathcal{Q}}_{i,j}\,, \displaybreak[0]\\[2ex]
\hat{\mathcal{Q}}_{i,j}&=\left(\mathcal{\hat{S}}\mathds{1}_4+\mathrm{i}\mathcal{\hat{P}}\gamma^{5}+\mathcal{\hat{V}}^{\mu}\gamma_{\mu}+\mathcal{\hat{A}}^{\mu}\gamma^{5}\gamma_{\mu}+\frac{1}{2}\mathcal{\hat{T}}^{\mu\nu}\sigma_{\mu\nu}\right)_{i,j}\,.
\end{align}
\end{subequations}

The indices stated explicitly are flavor indices, whereby spinor indices are suppressed, as usual. Writing the Dirac operator in this form turns out to be advantageous to apply a sophisticated algorithm for computing the determinant of a matrix in block form in a suggestive way \cite{Powell:2011nq}. The algorithm is a recursive method that will be briefly described as follows.

The base is to define $2N$ sets of matrices $\alpha^{(k)}_{i,j}$ with $k\in \{0\dots 2N-1\}$. Each set for a fixed $k$ contains $(2N)^2$ such matrices, i.e., we have $i$, $j\in \{1\dots 2N\}$. To avoid confusion, we emphasize again that all matrices are $(4\times 4)$, i.e., the indices are not spinor indices, but they label these matrices in flavor space. Now, the first step of the recursion is to assign the Dirac block at the position $(i,j)$ in flavor space to the matrix $\alpha^{(0)}_{i,j}$:
\begin{equation}
\label{eq:recursion-inicial-step}
\alpha^{(0)}_{i,j}\equiv \mathcal{D}_{i,j}\,.
\end{equation}

A recurrence relation is defined that allows for constructing new sets of matrices from previous ones. Hence, for $k\in \{0\dots 2N-2\}$ fixed, we obtain a new set of $(4\times 4)$ matrices $\{\alpha^{(k+1)}\}$ from the set $\{\alpha^{(k)}\}$ via
\begin{equation}
\label{eq:recursion-algorithm}
\alpha^{(k+1)}_{i,j}=\alpha^{(k)}_{i,j}-\alpha^{(k)}_{i,2N-k}(\alpha^{(k)}_{2N-k,2N-k})^{-1}\alpha^{(k)}_{2N-k,j}\,,
\end{equation}
where $A^{-1}$ denotes the inverse of the matrix $A$. Having these $2N$ sets of matrices at hand, the determinant of the original block matrix in \eqref{eq:dirac-operator-neutrinos} is given by
\begin{equation}
\label{eq:determinant-final-computation}
\det\mathcal{D}_{\upnu}=\prod_{k=1}^{2N} \det(\alpha^{(2N-k)}_{k,k})\,.
\end{equation}

Hence, to compute the determinant, only a subset of the matrices obtained before is necessary. This procedure has a great advantage compared to a brute-force evaluation of the determinant. In virtue of~\eqref{eq:determinant-final-computation}, the determinant of the full Dirac operator including the flavor structure decomposes into a product of determinants of matrizes that have the form of single-fermion Dirac operators.

Below, we intend to apply this powerful algorithm to the nonminimal neutrino sector. To simplify our notation, we introduce a generic Lorentz-violating operator $\hat{\mathcal{O}}^X$ with a suitable Lorentz index structure $X$. The latter can stand for one of the five possible operators: $\hat{\mathcal{O}}^X\in \{\hat{\mathcal{S}},\hat{\mathcal{P}},\hat{\mathcal{V}}^{\mu},\hat{\mathcal{A}}^{\mu},\hat{\mathcal{T}}^{\mu\nu}\}$. Now, the steps to be used for the algorithm are as follows:

\begin{enumerate}[leftmargin=*,labelsep=4.9mm]

\item Definition of initial operators:

According to \eqref{eq:recursion-inicial-step}, the first step of the recursion is
\begin{subequations}
\begin{equation}
\alpha^{(0)}_{i,j}=\left(\hat{\mathcal{S}}\mathds{1}_4+\mathrm{i}\hat{\mathcal{P}}\gamma^5+\hat{\mathcal{V}}^{\mu}\gamma_{\mu}+\hat{\mathcal{A}}^{\mu}\gamma^5\gamma_{\mu}+\frac{1}{2}\hat{\mathcal{T}}^{\mu\nu}\sigma_{\mu\nu}\right)^{(0)}_{i,j}\,,
\end{equation}
with
\begin{align}
\hat{\mathcal{S}}^{(0)}_{i,j}&=-M_{i,j}+\hat{\mathcal{S}}_{i,j}\,,\quad \hat{\mathcal{P}}^{(0)}_{i,j}=\hat{\mathcal{P}}_{i,j}\,,\quad \hat{\mathcal{V}}^{(0)\mu}_{i,j}=p^{\mu}\delta_{i,j}+\hat{\mathcal{V}}^{\mu}_{i,j}\,, \displaybreak[0]\\[2ex]
\hat{\mathcal{A}}^{(0)\mu}_{i,j}&=\hat{\mathcal{A}}^{\mu}_{i,j}\,,\quad \hat{\mathcal{T}}^{(0)\mu\nu}_{i,j}=\hat{\mathcal{T}}^{\mu\nu}_{i,j}\,.
\end{align}
\end{subequations}
\item Computation of inverse matrix:

The recurrence relation~(\ref{eq:recursion-algorithm}) requires the inverse of $\alpha^{(k)}_{i,j}$, which has the form of a single-fermion Dirac operator. Therefore, its inverse is linked to the fermion propagator obtained in \eqref{eq:propagator-single-fermion}~where
\begin{subequations}
\begin{equation}
(\alpha^{(k)}_{2N-k,2N-k})^{-1}=S|_{\hat{\mathcal{O}}^X=\hat{\mathcal{O}}^{(k)X}_{2N-k,2N-k}}\,.
\end{equation}

The denominator that appears in the latter expression is given by the left-hand side of the dispersion Equation~(\ref{eq:dispersion-equation-single-fermion}):
\begin{equation}
\Delta^{(k)}\equiv \det(\alpha^{(k)}_{2N-k,2N-k})=\Delta|_{\hat{\mathcal{O}}^X=\hat{\mathcal{O}}^{(k)X}_{2N-k,2N-k}}\,.
\end{equation}
\end{subequations}
\item Product of the second and third factor of recurrence relation:

Keeping the previous result in mind, the product necessary to evaluate \eqref{eq:recursion-algorithm} contains all kinds of combinations of Dirac matrices. As the 16 matrices of the set $\Gamma^A$ form a basis of $(4\times 4)$ matrices, these combinations can be completely expressed in terms of the 16 original matrices. To do so, a slew of matrix identities are indispensable, which are to be found in \appref{sec:relations-dirac-matrices}. The result of the product then has the following form:
\begin{align}
\label{eq:result-product-2-3}
(\alpha^{(k)}_{2N-k,2N-k})^{-1}\alpha^{(k)}_{2N-k,j}&=\frac{1}{\Delta^{(k)}}\left(\hat{\bar{\mathcal{S}}}+\mathrm{i}\hat{\bar{\mathcal{P}}}+\hat{\bar{\mathcal{V}}}^{\mu}\gamma_{\mu}+\hat{\bar{\mathcal{A}}}^{\mu}\gamma^5\gamma_{\mu}+\frac{1}{2}\hat{\bar{\mathcal{T}}}^{\mu\nu}\sigma_{\mu\nu}\right)_{2N-k,j}^{(k)}\,,
\end{align}
according to the single-fermion result of \eqref{eq:dispersion-equation-single-fermion}. A bar is added to the new operators that follow from this product. These are explicitly given by
\begin{subequations}
\label{eq:operators-bar}
\begin{align}
\hat{\bar{\mathcal{S}}}^{(k)}_{2N-k,j}&=\left(\hat{\mathcal{S}}^{(p)}\hat{\mathcal{S}}-\hat{\mathcal{P}}^{(p)}\hat{\mathcal{P}}+\hat{\mathcal{V}}^{(p)}\cdot\hat{\mathcal{V}}-\hat{\mathcal{A}}^{(p)}\cdot\hat{\mathcal{A}}-\frac{1}{2}\hat{\mathcal{T}}^{(p)}\cdot\hat{\mathcal{T}}\right)^{(k)}_{2N-k,j}\,, \displaybreak[0]\\[2ex]
\mathrm{i}\hat{\bar{\mathcal{P}}}^{(k)}_{2N-k,j}&=\left(\mathrm{i}\hat{\mathcal{S}}^{(p)}\hat{\mathcal{P}}+\mathrm{i}\hat{\mathcal{P}}^{(p)}\hat{\mathcal{S}}-\hat{\mathcal{V}}^{(p)}\cdot\hat{\mathcal{A}}+\hat{\mathcal{A}}^{(p)}\cdot\hat{\mathcal{V}}+\mathrm{i}[\star(\hat{\mathcal{T}}^{(p)}\wedge\hat{\mathcal{T}})]\right)^{(k)}_{2N-k,j}\,, \displaybreak[0]\\[2ex]
\label{eq:vector-operator-bar}
\hat{\bar{\mathcal{V}}}^{(k)\mu}_{2N-k,j}&=\left(\hat{\mathcal{S}}^{(p)}\hat{\mathcal{V}}^{\mu}+\hat{\mathcal{V}}^{(p)\mu}\hat{\mathcal{S}}+\mathrm{i}(\hat{\mathcal{V}}^{(p)}\cdot\hat{\mathcal{T}}+\hat{\mathcal{T}}^{(p)}\cdot\hat{\mathcal{V}})^{\mu}+\mathrm{i}(\hat{\mathcal{P}}^{(p)}\hat{\mathcal{A}}^{\mu}-\hat{\mathcal{A}}^{(p)\mu}\hat{\mathcal{P}})\right. \notag \\
&\phantom{{}={}}\hspace{0.25cm}\left.+[\star(\hat{\mathcal{A}}^{(p)}\wedge\hat{\mathcal{T}}+\hat{\mathcal{T}}^{(p)}\wedge\hat{\mathcal{A}})]^{\mu}\right)_{2N-k,j}^{(k)}\,, \displaybreak[0]\\[2ex]
\hat{\bar{\mathcal{A}}}^{(k)\mu}_{2N-k,j}&=\left(\hat{\mathcal{S}}^{(p)}\hat{\mathcal{A}}^{\mu}+\hat{\mathcal{A}}^{(p)\mu}\hat{\mathcal{S}}+\mathrm{i}(\hat{\mathcal{P}}^{(p)}\hat{\mathcal{V}}^{\mu}-\hat{\mathcal{V}}^{(p)\mu}\hat{\mathcal{P}})+\mathrm{i}(\hat{\mathcal{A}}^{(p)}\cdot\hat{\mathcal{T}}+\hat{\mathcal{T}}^{(p)}\cdot\hat{\mathcal{A}})^{\mu}\right. \notag \\
&\phantom{{}={}}\left.\hspace{0.2cm}+[\star(\hat{\mathcal{V}}^{(p)}\wedge\hat{\mathcal{T}}+\hat{\mathcal{T}}^{(p)}\wedge\hat{\mathcal{V}})]^{\mu}\right)^{(k)}_{2N-k,j}\,, \displaybreak[0]\\[2ex]
\frac{1}{2}\hat{\bar{\mathcal{T}}}^{(k)\mu\nu}_{2N-k,j}&=\left(\frac{1}{2}(\hat{\mathcal{S}}^{(p)}\hat{\mathcal{T}}^{\mu\nu}+\hat{\mathcal{T}}^{(p)\mu\nu}\hat{\mathcal{S}})-\frac{1}{2}(\hat{\mathcal{P}}^{(p)}\widetilde{\hat{\mathcal{T}}}{}^{\mu\nu}+\widetilde{\hat{\mathcal{T}}}{}^{(p)\mu\nu}\hat{\mathcal{P}})-\frac{\mathrm{i}}{2}\hat{\mathcal{V}}^{(p)\mu}\wedge\hat{\mathcal{V}}^{\nu}\right.
\notag \\
&\phantom{{}={}}\left.\hspace{0.2cm}+\,\frac{\mathrm{i}}{2}\hat{\mathcal{A}}^{(p)\mu}\wedge\hat{\mathcal{A}}^{\nu}+\frac{1}{2}[\star(\hat{\mathcal{A}}^{(p)}\wedge\hat{\mathcal{V}}-\hat{\mathcal{V}}^{(p)}\wedge\hat{\mathcal{A}})]^{\mu\nu}+\mathrm{i}(\hat{\mathcal{T}}^{(p)}\cdot\hat{\mathcal{T}})^{[\mu\nu]}\right)^{(k)}_{2N-k,j}\,,
\end{align}
\end{subequations}
where, for brevity, we omit the indices summed over. In the latter results, $\wedge$ stands for the exterior product (wedge product) of two tensors and $\star$ is the Hodge dual of a tensor (cf.~\appref{sec:definition-wedge-hodge} for a definition of these mathematical operations). We interpret the form of these expressions based on the example of the vector operator of \eqref{eq:vector-operator-bar}.

First, we will discuss the possible terms that occur. Only certain combinations of basic operators are permitted. In particular, a vector operator can be formed from combinations of the scalar $\hat{\mathcal{S}}$ and the vector $\hat{\mathcal{V}}$. Another possibility is to contract the vector $\hat{\mathcal{V}}$ with the tensor $\hat{\mathcal{T}}$. The fact that a combination of a pseudoscalar $\hat{\mathcal{P}}$ and a pseudovector $\hat{\mathcal{A}}$ transforms as a vector again, explains the third term. Finally, the pseudovector $\hat{\mathcal{A}}$ can be contracted with the tensor $\hat{\mathcal{T}}$ to form a pseudovector. The Hodge dual of the latter provides a vector.

Second, as one of the two operators of each term comes from the inverse (related to the propagator), each of the previously discussed possibilities appears twice. In the second possibility, the roles of the operators simply switch, i.e., what was the vector operator in the first possibility becomes the scalar operator and vice versa.
\item Product of the first factor in recurrence relation and the result of step (3):

Now we can evaluate the product of the three matrices in \eqref{eq:recursion-algorithm} completely, which~provides
\begin{equation}
\label{eq:result-product-1-2-3}
\alpha^{(k)}_{i,2N-k}(\alpha^{(k)}_{2N-k,2N-k})^{-1}\alpha^{(k)}_{2N-k,j}=\frac{1}{\Delta^{(k)}}\left(\hat{\bar{\bar{\mathcal{S}}}}\mathds{1}_4+\mathrm{i}\hat{\bar{\bar{\mathcal{P}}}}\gamma^5+\hat{\bar{\bar{\mathcal{V}}}}{}^{\mu}\gamma_{\mu}+\hat{\bar{\bar{\mathcal{A}}}}{}^{\mu}\gamma^5\gamma_{\mu}+\frac{1}{2}\hat{\bar{\bar{\mathcal{T}}}}{}^{\mu\nu}\sigma_{\mu\nu}\right)^{(k)}_{i,j}\,.
\end{equation}
Following the same procedure as before, we obtain a set of new operators indicated by a double bar. For example, the new vector operator is given by
\begin{align}
\label{eq:vector-operator-bar-bar}
\hat{\bar{\bar{\mathcal{V}}}}{}^{(k)\mu}_{i,j}&=\left(\hat{\mathcal{S}}\hat{\bar{\mathcal{V}}}^{\mu}+\hat{\mathcal{V}}^{\mu}\hat{\bar{\mathcal{S}}}+\mathrm{i}(\hat{\mathcal{V}}\cdot\hat{\bar{\mathcal{T}}}+\hat{\mathcal{T}}\cdot\hat{\bar{\mathcal{V}}})^{\mu}+\mathrm{i}(\hat{\mathcal{P}}\hat{\bar{\mathcal{A}}}{}^{\mu}-\hat{\mathcal{A}}{}^{\mu}\hat{\bar{\mathcal{P}}})\right. \notag \\
&\phantom{{}={}}\hspace{0.25cm}\left.+[\star(\hat{\mathcal{A}}\wedge\hat{\bar{\mathcal{T}}}+\hat{\mathcal{T}}\wedge\hat{\bar{\mathcal{A}}})]^{\mu}\right)_{i,j}^{(k)}\,.
\end{align}
We see that the structure of the latter result is the same as that of \eqref{eq:vector-operator-bar}. The simple difference is that the operators have to be renamed according to $\hat{\mathcal{O}}^{(p)X}\mapsto\hat{\mathcal{O}}^X$ and $\hat{\mathcal{O}}^X\mapsto \hat{\bar{\mathcal{O}}}{}^X$. Analog replacements must be performed for the remaining operators.
\item Recursive step $k\mapsto k+1$:

Now we have the ingredients to evaluate \eqref{eq:recursion-algorithm}:
\begin{subequations}
\begin{align}
\alpha^{(k+1)}_{i,j}&=\alpha^{(k)}_{i,j}-\alpha^{(k)}_{i,2N-k}(\alpha^{(k)}_{2N-k,2N-k})^{-1}\alpha^{(k)}_{2N-k,j} \notag \\
&=\left(\hat{\mathcal{S}}\mathds{1}_4+\mathrm{i}\hat{\mathcal{P}}\gamma^5+\hat{\mathcal{V}}^{\mu}\gamma_{\mu}+\hat{\mathcal{A}}^{\mu}\gamma^5\gamma_{\mu}+\frac{1}{2}\hat{\mathcal{T}}^{\mu\nu}\sigma_{\mu\nu}\right)^{(k+1)}_{i,j}\,.
\end{align}
Thus, the $(k+1)$-th operators are expressed in terms of the $k$-th operators via
\begin{equation}
\label{eq:generic-operator-recursive-step}
\hat{\mathcal{O}}^{(k+1)X}_{i,j}=\hat{\mathcal{O}}^{(k)X}_{i,j}-\frac{1}{\Delta^{(k)}}\hat{\bar{\bar{\mathcal{O}}}}{}^{(k)X}_{i,j}\,.
\end{equation}
\end{subequations}
\item Express final operators in terms of inicial ones:

We insert \eqref{eq:generic-operator-recursive-step} $k$ times into itself successively to obtain
\begin{equation}
\label{eq:generic-operator-recursive-sum}
\hat{\mathcal{O}}_{i,j}^{(k+1)X}=\hat{\mathcal{O}}_{i,j}^{(0)X}-\sum^k_{l=0} \frac{1}{\Delta^{(l)}}\hat{\bar{\bar{\mathcal{O}}}}{}^{(l)X}_{i,j}\,.
\end{equation}
\item Final computation of determinant:

All the previous results are employed to compute the determinant according to \eqref{eq:determinant-final-computation}. Doing so, it is reasonable to extract a product of denominators $\Delta^{(n)}$ from the expression such that the determinant itself is a polynomial instead of a sum of fractions of polynomials:
\begin{subequations}
\begin{equation}
\label{eq:determinant-final}
\det\mathcal{D}_{\upnu}=\prod^{2N}_{k=1} \det(\alpha^{(2N-k)}_{k,k})=\det(\alpha_{1,1}^{(2N-1)})\prod^{2N}_{k=2} \Delta^{(2N-k)}=\frac{\det(\tilde{\alpha}_{1,1}^{(2N-1)})}{(\Pi_{\Delta}^{(2N-2)})^3}\,,
\end{equation}
with
\begin{align}
\Pi^{(k)}_{\Delta}&\equiv\prod^k_{n=0} \Delta^{(n)}\,, \displaybreak[0]\\[2ex]
\tilde{\alpha}^{(2N-1)}_{1,1}&\equiv\Pi_{\Delta}^{(2N-2)}\alpha_{1,1}^{(2N-1)}\,.
\end{align}
\end{subequations}

In the forthcoming section, the final result will be stated explicitly.

\end{enumerate}

\section{Full Dispersion Equation of Neutrino Sector}
\label{sec:neutrino-dispersion-equation-result}

As the matrix $\tilde{\alpha}_{1,1}^{(2N-1)}$ contained in the final form of the determinant in \eqref{eq:determinant-final} has the same structure as the single-fermion Dirac theory in \eqref{eq:single-fermion-modified-dirac-theory}, we can directly compute the full dispersion equation for the nonminimal SME neutrino sector based on the Dirac operator of \eqref{eq:dirac-operator-modified-neutrinos}. The prefactors that we extracted from the determinant in \eqref{eq:determinant-final} do not play a role any longer. So neutrinos subject to any kind of Lorentz violation parameterized by the nonminimal SME obey the dispersion equation
\begin{subequations}
\begin{equation}
\label{eq:dispersion-equation-neutrinos}
0=(\breve{\mathcal{S}}_-^2-\breve{\mathcal{T}}_-^2)(\breve{\mathcal{S}}_+^2-\breve{\mathcal{T}}_+^2)+\breve{\mathcal{V}}^2_-\breve{\mathcal{V}}^2_+-2\breve{\mathcal{V}}_-\cdot(\breve{\mathcal{S}}_-\eta+2\mathrm{i}\breve{\mathcal{T}}_-)\cdot(\breve{\mathcal{S}}_+\eta-2\mathrm{i}\breve{\mathcal{T}}_+)\cdot\breve{\mathcal{V}}_+\,,
\end{equation}
with the operators
\begin{align}
\breve{\mathcal{S}}_{\pm}&=(\breve{\mathcal{S}}\pm\mathrm{i}\breve{\mathcal{P}})_{1,1}^{(2N-1)}\,, \displaybreak[0]\\[2ex]
\breve{\mathcal{V}}_{\pm}^{\mu}&=(\breve{\mathcal{V}}^{\mu}\pm\breve{\mathcal{A}}^{\mu})_{1,1}^{(2N-1)}\,, \displaybreak[0]\\[2ex]
\breve{\mathcal{T}}_{\pm}^{\mu\nu}&=\frac{1}{2}\left[\breve{\mathcal{T}}^{\mu\nu}\pm\mathrm{i}\widetilde{\breve{\mathcal{T}}}{}^{\mu\nu}\right]_{1,1}^{(2N-1)}\,,
\end{align}
and
\begin{equation}
\breve{\mathcal{O}}^{(2N-1)X}_{i,j}=\Pi_{\Delta}^{(2N-2)}\hat{\mathcal{O}}^{(2N-1)X}_{i,j}\,.
\end{equation}
\end{subequations}

The latter result is the central finding in the current paper. The interesting observation is that the dispersion equation for $N$ neutrino flavors (including the description of both Dirac and Majorana neutrinos) subject to Lorentz violation has a form completely analog to the dispersion equation of the single-fermion sector stated in \eqref{eq:dispersion-equation-single-fermion}. In general, the coefficients appearing in the dispersion equation are (lengthy) combinations of the neutrino coefficients that are given by a subsequent application of Equations~(\ref{eq:operators-bar}) and  (\ref{eq:vector-operator-bar-bar}) (for the vector operator, in particular) and \eqref{eq:generic-operator-recursive-step}. This procedure has to be repeated a sufficient number of times to be able to compute the final necessary operator via \eqref{eq:generic-operator-recursive-sum}. The advantage of the result given by
Equation (\ref{eq:dispersion-equation-neutrinos}) is that it is covariant and does not apply to only a specific observer frame. We think that this form is also suitable to be used in a computer algebra system.

\section{First-Order Behavior of Dispersion Relations}
\label{sec:first-order-dispersion-relations}

As \eqref{eq:dispersion-equation-neutrinos} is, in general, a polynomial in $p_0$ of high degree, it is challenging to obtain exact dispersion relations from it. Thus, in the current section we intend to get some idea on the general structure of modified neutrino dispersion relations at leading order in Lorentz violation. Since $\hat{\mathcal{S}}_{i,j}^{(k)}$ and $\hat{\mathcal{V}}_{i,j}^{(k)\mu}$ exhibit Lorentz-invariant parts, it is reasonable to decompose the latter operators into two contributions to separate both pieces from each other. By doing so, we get
\begin{subequations}
\begin{equation}
\label{eq:operator-first-order-1}
\hat{\mathcal{O}}_{i,j}^{(k)}=\hat{\mathcal{O}}_{i,j}^{(0)}-\sum_{l=0}^{k-1}\frac{1}{\Delta^{(l)}_{2N-l,2N-l}}\hat{\bar{\bar{\mathcal{O}}}}^{(l)}_{i,j}\approx \hat{\mathcal{O}}^{(k)}_{0;i,j}+\delta\hat{\mathcal{O}}^{(k)}_{i,j}\,.
\end{equation}

The additional index ``0'' (without parentheses) denotes a Lorentz-invariant part. In an analog manner, we generically expand the operators $\hat{\mathcal{P}}_{i,j}^{(k)}$, $\hat{\mathcal{A}}_{i,j}^{(k)\mu}$ and $\hat{\mathcal{T}}_{i,j}^{(k)\mu\nu}$ in the form
\begin{equation}
\label{eq:operator-first-order-2}
\hat{\mathcal{O}}_{i,j}^{(k)}=\hat{\mathcal{O}}^{(0)}_{i,j}-\sum_{l=0}^{k-1}\frac{1}{\Delta^{(l)}_{2N-l,2N-l}}\hat{\bar{\bar{\mathcal{O}}}}^{(l)}_{i,j}\approx\delta\hat{\mathcal{O}}^{(k)}_{i,j}\,.
\end{equation}
\end{subequations}

Here, the notation $\delta$ indicates that all leading-order Lorentz-violating contributions are included (recall that $\hat{\mathcal{P}}$, $\hat{\mathcal{A}}$ and $\hat{\mathcal{T}}$ must be expanded to second order). Later on we can substitute those parts by the corresponding expansions. The expanded dispersion equation then takes the form
\begin{subequations}
\begin{align}
0&\approx \left[(\hat{\mathcal{S}}_{0;i,j}^{(k)})^2+2\hat{\mathcal{S}}_{0;i,j}^{(k)}\delta\hat{\mathcal{S}}_{i,j}^{(k)}-(\hat{\mathcal{V}}_{0;i,j}^{(k)})^2-2\hat{\mathcal{V}}_{0;i,j}^{(k)}\cdot\delta\hat{\mathcal{V}}_{i,j}^{(k)}\right]^2 \notag \\
&\phantom{{}={}}+2\left[(\hat{\mathcal{S}}_{0;i,j}^{(k)})^2-(\hat{\mathcal{V}}_{0;i,j}^{(k)})^2\right](\delta\hat{\mathcal{P}}_{i,j}^{(k)})^2-4(\delta\hat{\mathcal{Y}}_{i,j}^{(k)})^2\,,
\end{align}
with the spin-nondegenerate part
\begin{align}
\label{eq:spin-nondegenerate-first-order}
(\delta\hat{\mathcal{Y}}_{i,j}^{(k)})^2&=(\hat{\mathcal{V}}_{0;i,j}^{(k)}\cdot \delta\hat{\mathcal{A}}_{i,j}^{(k)})^2-\frac{1}{2}\left[(\hat{\mathcal{V}}_{0;i,j}^{(k)})^2+(\hat{\mathcal{S}}_{0;i,j}^{(k)})^2\right](\delta\hat{\mathcal{A}}_{i,j}^{(k)})^2+2\hat{\mathcal{S}}_{0;i,j}^{(k)}\hat{\mathcal{V}}_{0;i,j}^{(k)}\cdot\delta\hat{\mathcal{\tilde{T}}}_{i,j}^{(k)}\cdot \delta\hat{\mathcal{A}}_{i,j}^{(k)} \notag \\
&\phantom{{}={}}+\hat{\mathcal{V}}_{0;i,j}^{(k)}\cdot \delta\hat{\mathcal{\tilde{T}}}_{i,j}^{(k)}\cdot \delta\hat{\mathcal{\tilde{T}}}_{i,j}^{(k)}\cdot\hat{\mathcal{V}}_{0;i,j}^{(k)}+\frac{1}{4}\left[(\hat{\mathcal{V}}_{0;i,j}^{(k)})^2-(\hat{\mathcal{S}}_{0;i,j}^{(k)})^2\right](\delta\hat{\mathcal{\tilde{T}}}_{i,j}^{(k)})^2\,.
\end{align}
\end{subequations}

Note that the last term on the right-hand side of the latter equation vanishes for the single-fermion sector. The leading-order expansion of the dispersion equation can be further expressed as
\begin{align}
(\hat{\mathcal{V}}_{0;i,j}^{(k)})^2-(\hat{\mathcal{S}}_{0;i,j}^{(k)})^2&\approx 2\hat{\mathcal{S}}_{0;i,j}^{(k)}\delta\hat{\mathcal{S}}_{i,j}^{(k)}-2\hat{\mathcal{V}}_{0;i,j}^{(k)}\cdot \delta\hat{\mathcal{V}}_{i,j}^{(k)} \notag \\
&\phantom{{}={}}\pm 2\sqrt{(\delta\hat{\mathcal{Y}}_{i,j}^{(k)})^2-\frac{1}{2}\left[(\hat{\mathcal{S}}_{0;i,j}^{(k)})^2-(\hat{\mathcal{V}}_{0;i,j}^{(k)})^2\right](\delta\hat{\mathcal{P}}_{i,j}^{(k)})^2}\,.
\end{align}

It shall be emphasized again that the Lorentz-invariant pieces are contained only in $\hat{\mathcal{V}}_{0;i,j}^{(k)}$ and $\hat{\mathcal{S}}_{0;i,j}^{(k)}$, respectively. The term on the left-hand side of the equation above is a polynomial of order $8N$ in $p^0$ and contains the pure-mass part. The standard dispersion relations for the case of an arbitrary number of $N$ flavors seem to follow the pattern
\begin{subequations}
\begin{equation}
\label{eq:standard-dispersion-relation-generic}
E_0^{(u)}\overset{?}{=}\sqrt{\mathbf{p}^2+\frac{1}{2N}\left[\mathfrak{M}+(m_{\mathrm{eff}}^{(u)})^2\right]}\,,\quad u\in \{1\dots 2N\}\,,
\end{equation}
where $\mathbf{p}$ is the spatial momentum of $p^{\mu}$ and
\begin{equation}
\mathfrak{M}=\mathrm{Tr}(M^2)=\sum_{i,j=1}^{2N}M_{i,j}M_{j,i}\,.
\end{equation}
\end{subequations}

Furthermore, $(m_{\mathrm{eff}}^{(u)})^2$ for $u$ fixed is a complicated polynomial of mass matrix coefficients that we simply called an ``effective mass squared.'' The validity of \eqref{eq:standard-dispersion-relation-generic} is challenging to proof for an arbitrary $N$, though. Taking Lorentz violation into account, the first-order modification is given by
\begin{align}
E^{(u)}&\approx E_0^{(u)}+\frac{1}{2NE_0^{(u)}(m_{\mathrm{eff}}^{(u)})^2}\left\{\hat{\mathcal{S}}_{0;i,j}^{(k)}\delta\hat{\mathcal{S}}_{i,j}^{(k)}-\hat{\mathcal{V}}_{0;i,j}^{(k)}\cdot\delta\hat{\mathcal{V}}_{i,j}^{(k)}\right. \notag\\
&\phantom{{}={}}\hspace{3.7cm}\left.\pm \sqrt{(\delta\hat{\mathcal{Y}}_{i,j}^{(k)})^2-\frac{1}{2}\left[(\hat{\mathcal{S}}_{0;i,j}^{(k)})^2-(\hat{\mathcal{V}}_{0;i,j}^{(k)})^2\right] (\delta\hat{\mathcal{P}}_{i,j}^{(k)})^2}\,\right\}\,,
\end{align}
with the operators defined in Equations~(\ref{eq:operator-first-order-1}), (\ref{eq:operator-first-order-2}) and (\ref{eq:spin-nondegenerate-first-order}). Here we see how the presence of the spin-nondegenerate operators doubles the number of dispersion relations, as expected. In total, there are then $8N$ modified dispersion laws, which corresponds to the degree of the polynomial in $p^0$. Furthermore, the pseudoscalar operator also leads to such a doubling. If the pseudo-scalar operator is the only source for Lorentz violation, it contributes at second order, as $\delta\hat{\mathcal{P}}_{i,j}^{(k)}$ is of second order in Lorentz violation.

\subsection{Special Case: $N=1$}

As even the general first-order expansion is quite complicated, it shall be exemplified as follows. We consider the theory of a single neutrino flavor that can be of either Dirac or Majorana type. It is reasonable to switch Lorentz violation off at first. The dispersion equation is then a polynomial of fourth degree. (In principle, the polynomial on the right-hand side of the dispersion equation is raised to the second power, that is, the degeneracy of all zeros is doubled.) It reads:
\begin{subequations}
\begin{equation}
0=p^4-\mathfrak{M}p^2+\tilde{\mathfrak{M}}\,,
\end{equation}
where
\begin{equation}
\label{eq:matrix-coefficients-quantities}
\mathfrak{M}=\mathrm{Tr}(M^2)\,,\quad \tilde{\mathfrak{M}}=(\det M)^2\,.
\end{equation}
\end{subequations}

Solving for $p_0$ delivers two distinct energies:
\begin{equation}
E_0^{(1,2)}=\sqrt{\mathbf{p}^2+\frac{\mathfrak{M}}{2}\pm\sqrt{\left(\frac{\mathfrak{M}}{2}\right)^2-\tilde{\mathfrak{M}}}}\,.
\end{equation}

Now, the neutrino energies modified by Lorentz violation are found to have the form
\begin{subequations}
\label{eq:operators-first-order-dispersion-relations}
\begin{equation}
\label{eq:dispersion-relations-N1}
E^{(1,2)\pm}=E_0^{(1,2)}-\frac{1}{2E_0^{(1,2)}}\left(\hat{\mathcal{S}}^{(1,2)}_{\mathrm{disp}}+p\cdot\hat{\mathcal{V}}^{(1,2)}_{\mathrm{disp}}\pm\delta\hat{\mathcal{Y}}\right)+\dots\,,
\end{equation}
with
\begin{align}
\label{eq:scalar-operator-dispersion-relation}
\hat{\mathcal{S}}^{(1,2)}_{\mathrm{disp}}&=M_{ab}\hat{\mathcal{S}}_{ba}\pm \frac{(M_{aa}+M_{bb})M_{ab}\hat{\mathcal{S}}_{ba}-(M_{aa}^2+M_{aa}M_{bb}-2M_{ab}M_{ba})\hat{\mathcal{S}}_{aa}}{\sqrt{(M_{11}-M_{22})^2+4M_{12}M_{21}}}\,, \displaybreak[0]\\[2ex]
\label{eq:vector-operator-dispersion-relation}
\hat{\mathcal{V}}^{(1,2)\mu}_{\mathrm{disp}}&=\hat{\mathcal{V}}^{\mu}_{aa}\pm\frac{2M_{ab}\hat{\mathcal{V}}^{\mu}_{ba}-M_{aa}\hat{\mathcal{V}}^{\mu}_{bb}}{\sqrt{(M_{11}-M_{22})^2+4M_{12}M_{21}}}\,, \displaybreak[0]\\[2ex]
(\delta\hat{\mathcal{Y}})^2&=\frac{1}{(M_{11}+M_{22})^{2}\left[(M_{11}-M_{22})^2+4M_{12}M_{21}\right]} \notag \\
&\phantom{{}={}}\times\left\{(\hat{\mathcal{V}}_{0}\cdot \delta\hat{\mathcal{A}})^2-\frac{1}{2}\left[(\hat{\mathcal{V}}_0)^{2}+(\hat{\mathcal{S}}_0)^{2}\right](\delta\hat{\mathcal{A}})^2+2\hat{\mathcal{S}}_0\hat{\mathcal{V}}_0\cdot\delta
\hat{\mathcal{\tilde{T}}}\cdot\delta\hat{\mathcal{A}}\right. \notag \\
&\phantom{{}={}}\hspace{0.7cm}\left.+\hat{\mathcal{V}}_0\cdot\delta\hat{\mathcal{\tilde{T}}}\cdot \delta\hat{\mathcal{\tilde{T}}}\cdot\hat{\mathcal{V}}_0+\frac{1}{4}\left[(\hat{\mathcal{V}}_0)^2-(\hat{\mathcal{S}}_0)^2\right] (\delta\hat{\mathcal{\tilde{T}}})^2\right\}_{1,1}^{(1)}\,.
\end{align}
\end{subequations}

Furthermore,
\begin{subequations}
\begin{align}
\delta\hat{\mathcal{A}}_{1,1}^{(1)\mu}&\approx \hat{\mathcal{A}}_{1,1}^{(1)\mu}=\hat{\mathcal{A}}_{1,1}^{(0)\mu}-\frac{1}{\Delta_{2,2}^{(0)}}\hat{\bar{\bar{\mathcal{A}}}}_{1,1}^{(0)\mu}\,, \displaybreak[0]\\[2ex]
\delta \hat{\mathcal{T}}_{1,1}^{(k)\mu\nu}&\approx \hat{\mathcal{T}}_{1,1}^{(1)\mu\nu}=\hat{\mathcal{T}}_{1,1}^{(0)\mu\nu}-\frac{1}{\Delta_{2,2}^{(0)}}\hat{\bar{\bar{\mathcal{T}}}}_{1,1}^{(0)\mu\nu}\,,
\end{align}
and
\begin{align}
\hat{\mathcal{S}}_0&=-M_{11}-\frac{M_{12}M_{22}M_{21}}{p^2-M_{22}}\,, \displaybreak[0]\\
\hat{\mathcal{V}}_0^{\mu}&=\frac{p^2-M_{12}M_{21}-M_{11}^2}{p^2-M_{22}}p^{\mu}\,.
\end{align}
\end{subequations}

The flavor indices in Equations~(\ref{eq:scalar-operator-dispersion-relation}) and  (\ref{eq:vector-operator-dispersion-relation}) are understood to be summed over.
Note that in~\eqref{eq:dispersion-relations-N1} two signs can be chosen at different positions independently from each other. The first sign is indicated by the suffices (1,2) and appears in the Lorentz-invariant and spin-degenerate parts of the dispersion relation. The second sign is marked by the additional index $\pm$ and is related to the spin-nondegenerate coefficients only, which are incorporated in the quantity $(\delta\hat{\mathcal{Y}})^2$. Hence, for $N=1$ there can be already 4 different dispersion relations. With spin-nondegenerate Lorentz violation present, there are two distinct dispersion relations for each of the two neutrino types. The number of modified dispersion laws is supposed to increase with the number of neutrino flavors.

\section{Classical Lagrangians}
\label{sec:classical-lagrangians}

As a final application of all the previous results, we want to map the neutrino field theory to a theory of classical, relativistic, pointlike particles. The latter is described by a Lagrange function in terms of the four-velocity $u^{\mu}$. The technique to carry out such a mapping in the context of the SME was developed in Reference \cite{Kostelecky:2010hs} and has been subject to intense studies for the past ten years. Presently, the procedure is well-known and has been applied to both the minimal and nonminimal SME fermion sector. At leading order in Lorentz violation, it was demonstrated that a classical Lagrangian can be obtained from the dispersion relation directly via a mapping procedure \cite{Reis:2017ayl}. We will employ this method for the case $N=1$, which leads to the classical Lagrangians corresponding to this field theory of modified neutrinos. The Lagrangians can be cast into the form
\begin{subequations}
\label{eq:classical-lagrangian-neutrino}
\begin{align}
L^{(1,2)\pm}&=-M^{(1,2)}\sqrt{u^2}\left[1-\frac{(\hat{\mathcal{S}}_{\mathrm{disp}}^{(1,2)})_{\ast}\pm\delta\hat{\mathcal{Y}}_{\ast}}{2(M^{(1,2)})^2}-\frac{u\cdot(\hat{\mathcal{V}}_{\mathrm{disp}}^{(1,2)})_{\ast}}{2M^{(1,2)}\sqrt{u^2}}+\dots\right]\,, \displaybreak[0]\\[2ex]
M^{(1,2)}&=\sqrt{\frac{\mathfrak{M}}{2}\pm\sqrt{\left(\frac{\mathfrak{M}}{2}\right)^2-\tilde{\mathfrak{M}}}}\,,
\end{align}
\end{subequations}
with $\mathfrak{M}$ and $\tilde{\mathfrak{M}}$ of \eqref{eq:matrix-coefficients-quantities}. The quantities endowed with an asterisk emerge from the expressions defined in \eqref{eq:operators-first-order-dispersion-relations} in replacing each Lorentz-violating operator by a suitable contraction of the corresponding controlling coefficients with four-velocities. Each contraction involves an additional prefactor that depends on the mass dimension. For a generic Lorentz-violating operator of mass dimension $d$, this contraction reads
\begin{equation}
(\hat{\mathcal{O}}_{i,j}^{(k)})_{\ast}=\left(\frac{M^{(1,2)}}{\sqrt{u^2}}\right)^{d-3}\mathcal{O}_{i,j}^{(k)\alpha_1\dots\alpha_{d-3}}u_{\alpha_1}\dots u_{\alpha_{d-3}}\,.
\end{equation}

As there are four modified dispersion relations, there are also four classical Lagrangians. Each of those describes a type of massive ``classical neutrino.'' For vanishing Lorentz violation, the Lagrangians take the form $L^{(1,2)}=-M^{(1,2)}\sqrt{u^2}$, which is analogous to the standard result $L=-m_{\psi}\sqrt{u^2}$ for a single Dirac fermion of mass $m_{\psi}$ under the identification $m_{\psi}=M^{(1,2)}$. Here we see how the combinations $M^{(1,2)}$ of mass matrix coefficients can be interpreted as something like a simple mass of the classical neutrino analog. This behavior is a classical remnant of the quantum effect of neutrino mixing. Furthermore, the classical Lagrangian can be checked to be positively homogeneous of degree 1 in $u^{\mu}$, as expected.

Classical Lagrangians as those of Equation~(\ref{eq:classical-lagrangian-neutrino}) are valuable in the description of Lorentz violation for neutrinos in the presence of an external gravitational field. Since the current section serves only as a demonstration of the basic procedure, we will not delve deeper into this interesting topic. Whenever the neutrino masses are simply neglected, such classical Lagrangians simply lose their meaning. Neutrino propagation in a gravitational background must then be described with a different formalism such as the eikonal equation, which turned out to be fruitful for massless particles, e.g., photons~\cite{Schreck:2015dsa}.

\section{Conclusions}
\label{sec:conclusions}

In this paper we derived the full propagator of a single-fermion Dirac theory based on the nonminimal SME as well as the full dispersion equation of modified neutrinos. Both results are expressed in a covariant form. Although it is quite an essential tool in perturbation theory, the full propagator of the nonminimal SME fermion sector has not been stated elsewhere, so far. The dispersion equation is valid for the general case of $N$ neutrino flavors with the description of both Dirac and Majorana neutrinos included. Despite the additional flavor structure and the distinction between Dirac and Majorana neutrinos, we found that the dispersion equation has a structure analogous to that of a single Dirac fermion. However, it is also clear that the form of the Lorentz-violating operators that occur in the neutrino dispersion equation is much more involved because of the additional flavor structure. We also investigated the dispersion relations at leading order in Lorentz violation for basic configurations. Finally, we included a brief comment on classical Lagrangians in the context of neutrinos. Our findings are technical, but they may be valuable in forthcoming phenomenological works on Lorentz-violating Dirac fermions and neutrinos.

\authorcontributions{Formal analysis, J.A.A.S and M.S.; writing--original draft preparation, M.S.; writing--review and editing, J.A.A.S and M.S.}

\funding{This research was funded by CNPq grant numbers Universal 421566/2016-7,  Produtividade 312201/2018-4 and by FAPEMA grant number Universal 01149/17.}

\acknowledgments{The authors are grateful for financial support by the Brazilian agencies CNPq, CAPES, and FAPEMA. We thank the two anonymous referees for a number of constructive comments on the first submitted version of the manuscript.}

\conflictsofinterest{The authors declare no conflict of interest.} 

\appendixtitles{yes} 
\appendix

\section{Useful Relations for Dirac Matrices}
\label{sec:relations-dirac-matrices}

For future reference, we think that it is a good idea to list the relations for Dirac matrices that we used to obtain our results. First, it is difficult to find the whole set of relations at a particular place. Second, in some works we even encountered typos. Therefore, the validity of the following relations was checked explicitly and they are supposed to be correct, as they stand:
\begin{subequations}
\begin{align}
\gamma_{\mu}\gamma_{\nu}&=\eta_{\mu\nu}\mathds{1}_{4}-\mathrm{i}\sigma_{\mu\nu}\,, \displaybreak[0]\\[2ex]
\sigma_{\mu\nu}\gamma^{5}&=\frac{\mathrm{i}}{2}\varepsilon_{\alpha\beta\mu\nu}\sigma^{\alpha\beta}\,, \displaybreak[0]\\[2ex]
\gamma_{\mu}\gamma_{\nu}\gamma^{5}&=\eta_{\mu\nu}\gamma^5+\frac{1}{2}\varepsilon_{\mu\nu\alpha\beta}\sigma^{\alpha\beta}\,, \displaybreak[0]\\[2ex]
\gamma_{\mu}\gamma_{\nu}\gamma_{\lambda}&=\eta_{\mu\nu}\gamma_{\lambda}+\eta_{\nu\lambda}\gamma_{\mu}-\eta_{\lambda\mu}\gamma_{\nu}+\mathrm{i}\varepsilon_{\mu\nu\lambda\rho}\gamma^{\rho}\gamma^{5}\,, \displaybreak[0]\\[2ex]
\sigma_{\mu\nu}\gamma_{\lambda}&=\mathrm{i}(\eta_{\nu\lambda}\gamma_{\mu}-\eta_{\lambda\mu}\gamma_{\nu})-\varepsilon_{\mu\nu\lambda\rho}\gamma^{\rho}\gamma^{5}\,, \displaybreak[0]\\[2ex]
\gamma_{\mu}\sigma_{\nu\lambda}&=\mathrm{i}(\eta_{\mu\nu}\gamma_{\lambda}-\eta_{\lambda\mu}\gamma_{\nu})-\varepsilon_{\mu\nu\lambda\rho}\gamma^{\rho}\gamma^{5}\,, \displaybreak[0]\\[2ex]
\gamma^{5}\sigma_{\mu\nu}\gamma_{\lambda}&=\mathrm{i}(\eta_{\nu\lambda}\gamma^5\gamma_{\mu}-\eta_{\lambda\mu}\gamma^5\gamma_{\nu})+\varepsilon_{\mu\nu\lambda\rho}\gamma^{\rho}\,, \displaybreak[0]\\[2ex]
\gamma^{5}\gamma_{\mu}\sigma_{\nu\lambda}&=\mathrm{i}(\eta_{\mu\nu}\gamma^5\gamma_{\lambda}-\eta_{\lambda\mu}\gamma^5\gamma_{\nu})+\varepsilon_{\mu\nu\lambda\rho}\gamma^{\rho}\,, \displaybreak[0]\\[2ex]
[\sigma_{\mu\nu},\sigma_{\lambda\alpha}]&=2\mathrm{i}(\eta_{\mu\lambda}\sigma_{\alpha\nu}-\eta_{\nu\lambda}\sigma_{\alpha\mu}+\eta_{\nu\alpha}\sigma_{\lambda\mu}-\eta_{\mu\alpha}\sigma_{\lambda\nu})\,, \displaybreak[0]\\[2ex]
\{\sigma_{\mu\nu},\sigma_{\lambda\alpha}\}&=2\left[(\eta_{\alpha\nu}\eta_{\lambda\mu}-\eta_{\alpha\mu}\eta_{\lambda\nu})\mathds{1}_4+\mathrm{i}\varepsilon_{\mu\nu\lambda\alpha}\gamma^5\right]\,, \displaybreak[0]\\[2ex]
\sigma_{\mu\nu}\sigma_{\lambda\alpha}&=\mathrm{i}(\eta_{\mu\lambda}\sigma_{\alpha\nu}-\eta_{\nu\lambda}\sigma_{\alpha\mu}+\eta_{\nu\alpha}\sigma_{\lambda\mu}-\eta_{\mu\alpha}\sigma_{\lambda\nu})+(\eta_{\alpha\nu}\eta_{\lambda\mu}-\eta_{\alpha\mu}\eta_{\lambda\nu})\mathds{1}_{4}+\mathrm{i}\varepsilon_{\mu\nu\lambda\alpha}\gamma^{5}\,.
\end{align}
\end{subequations}

We employed these results mainly to obtain Equations~(\ref{eq:result-product-2-3}) and  (\ref{eq:result-product-1-2-3}).

\section{Definition of Wedge Product and Hodge Dual}
\label{sec:definition-wedge-hodge}

The wedge product and Hodge dual are concepts that are of wide use in algebra. They turned out to be fruitful to express \eqref{eq:operators-bar} in a relatively compact form. We define the wedge product of two (contravariant) tensors $A$ and $B$ as the antisymmetrized direct product of these tensors:
\begin{equation}
\frac{N_a!N_b!}{(N_a+N_b)!}A^{\mu\nu\dots}\wedge B^{\alpha\beta\dots}\equiv A^{[\mu\nu\dots}B^{\alpha\beta\dots]}\equiv C^{\mu\nu\dots\alpha\beta\dots}\,,
\end{equation}
where $N_{a,b}$ is the number of indices of the tensor $A$ and $B$, respectively. The latter product gives rise to a new tensor $C$ with the union of Lorentz indices of the tensors $A$ and $B$. The definition of the Hodge dual of a tensor depends on the number of dimensions of the space considered. As we work in four-dimensional Minkowski spacetime, the Hodge dual of a covariant two-tensor $A$ provides a contravariant two tensor:
\begin{equation}
(\star A)^{\alpha\beta}\equiv \frac{1}{2}A_{\mu\nu}\varepsilon^{\mu\nu\alpha\beta}\,.
\end{equation}

On the other hand, the Hodge dual of a covariant three-tensor $B$ gives a contravariant vector and that of a four-tensor $C$ is a scalar:
\begin{equation}
(\star B)^{\varrho}\equiv \frac{1}{3!}B_{\mu\nu\lambda}\varepsilon^{\mu\nu\lambda\varrho}\,,\quad \star C\equiv \frac{1}{4!}C_{\mu\nu\lambda\alpha}\varepsilon^{\mu\nu\lambda\alpha}\,.
\end{equation}

Note that we follow the convention of the prefactors used in mathematics.

%



\reftitle{References}





\end{document}